\documentclass[prl,showpacs,preprintnumbers,amsmath,amssymb,twocolumn]{revtex4}

\usepackage{graphicx}
\usepackage{dcolumn}
\usepackage{bm}

\preprint{submitted to Journal of Spintronics and Magnetic
Nanomaterials - Special Issue}

\begin{document}

\title{Slater-Pauling behavior in half-metallic magnets}

\author{Iosif Galanakis}

\affiliation{Department of Materials Science, School of Natural
Sciences, University of Patras, GR-26504 Patra, Greece}

\date{\today}

\begin{abstract}
We review the appearance of Slater-Pauling rules in half-metallic
magnets. These rules have been derived using ab-initio electronic
structure calculations and directly connect the electronic
properties (existence of minority-spin energy gap) to the magnetic
properties (total spin magnetic moment) in these compounds. Their
exact formulation depends on the half-metallic family under study
and they can be easily derived if the hybridization of the
orbitals at various sites is taken into account.
\ \\ \textbf{Corresponding author :
} \underline{Iosif Galanakis} \textit{Postal address: Department
of Materials Science,
 University of Patras, GR-26504 Patra, Tel.:+30-2610-969925, Fax:+30-2610-969925, E-mail address:
galanakis@upatras.gr}
\end{abstract}

\maketitle

\section{Introduction}\label{sec1}

The developments in electronics, combining the magnetic and
semiconducting materials (so-called magnetoelectronis or
spintronics) \cite{Zutic2004,Felser,Zabel}, have brought
half-metallic magnets, initially predicted by de Groot and
collaborators in 1983 \cite{deGroot}, to the center of scientific
research. In these materials the two spin bands show a completely
different behavior. While the spin-up electronic band structure is
metallic, in the spin-down band the Fermi level falls within an
energy gap as in semiconductors \cite{Katsnelson}. Such
half-metallic compounds exhibit, ideally, a 100\% spin
polarization at the Fermi level and therefore they should have a
fully spin-polarized current and be ideal spin injectors into a
semiconductor, thus maximizing the efficiency of spintronic
devices \cite{Wolf}.

\begin{figure}
\begin{center}
\includegraphics[width=\columnwidth]{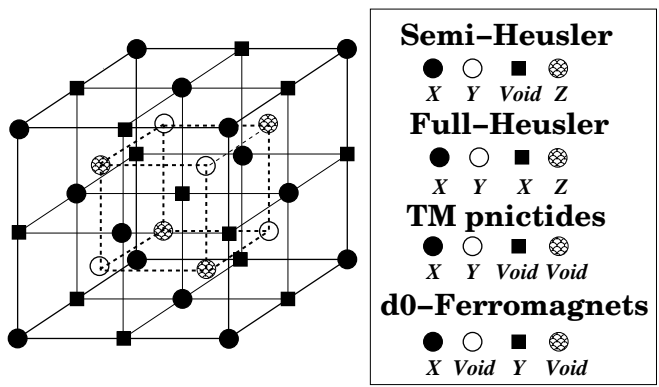}
\end{center}
\caption{Schematic representation of (i) the C1$_b$ cubic
structure adopted by the semi-Heusler compounds, (ii) the L2$_1$
structure adopted by the full-Heusler compounds, (iii) the
zinc-blende structure adopted by the transition-metal pnictides
and chalcogenides, and (iv) the rocksalt structure of the
d$^0$-ferromagnets. The cube contains exactly four unit cells.}
\label{fig1}
\end{figure}

The interest on half-metallic magnets has been mainly focused on
the Heusler compounds. The are two distinct families of Heuslers.
The so-called full-Heuslers, with the chemical formula X$_2$YZ,
crystallize in the L2$_1$ structure which consists of four fcc
sublattices, where X is a high valent transition or noble metal
atom, Y is a low-valent transition metal atom and Z is a sp
element \cite{landolt,landolt2}. In Fig. \ref{fig1} we present the
lattice structures of all the compounds revised in this Short
Review.
 The second class of Heuslers encompass the so-called semi- (or half-)
 Heusler compounds of
 the chemical form XYZ, crystallizing in the C1$_b$ structure. The structure
 as shown in Fig. \ref{fig1} is similar
 to the L2$_1$ one but now one of the four sites along the diagonal is empty.
  NiMnSb, a semi-Heusler, was the first one predicted to be a half-metal
  in 1983 by  de Groot and collaborators \cite{deGroot}.
The main advantages of Heusler alloys with respect to other
half-metallic systems (e.g. some oxides like CrO$_2$ and
Fe$_3$O$_4$ and some manganites like La$_{0.7}$Sr$_{0.3}$MnO$_3$)
\cite{Soulen98}) are their relatively high Curie temperatures
$T_C$ \cite{landolt,landolt2} (while for the other compounds $T_C$
is near the room temperature,  \textit{e.g.} for NiMnSb it is 730
K and for Co$_2$MnSi it reaches the 985 K \cite{landolt}) as well
as their structural similarity to the zincblende structure,
adopted by binary semiconductors widely used in industry (such as
GaAs on ZnS).

In 2000, Akinaga and collaborators \cite{Akinaga00} reported
growth of a few layers of CrAs in the metastable zincblende phase
(zb) for the first time; this was achieved by molecular beam
epitaxy on a GaAs substrate (the ground state structure of CrAs is
the MnP structure).  In the zb phase, CrAs was reported to be
ferromagnetic, with a Curie point higher than 400~K. In the same
publication, first-principles calculations of bulk zinc-blende
CrAs showed the material to be a half-metallic ferromagnet; the
moment per formula unit was calculated to be 3~$\mu_B$, in
agreement with the experimental result. These findings initiated a
strong activity on the transition metal (TM) pnictides and
chalcogenides, because several merits were combined as in Heusler
compounds mentioned above: the half-metallic property, coherent
growth on a semiconductor, and $T_C$ higher than room temperature.
The activity was extended beyond zb CrAs, encompassing a variety
of tetrahedrally bonded TM compounds with $sp$ atoms of the IV, V
and VI groups of the periodic table \cite{ReviewCrAs}.

Except the TM pnictides and chalcogenides, the half-metallic
property was also predicted in another class of binary materials
the so-called sp-electron ferromagnets or d$^0$-ferromagnets
\cite{Volniaska10}. In two pioneering papers published by Geshi et
al \cite{Geshi04} and Kusakabe et al \cite{Kusakabe04}, it was
shown
 using first-principles calculations that CaP, CaAs and CaSb
compounds present half-metallic ferromagnetism when grown in the
zincblende structure. Several first-principles calculations have
been carried in several  I/II-IV/V compounds and it was shown that
they are half-metallic magnets in all three  zincblende, wurtzite
and rocksalt (rs) metastable structures; the ground state in all
cases being the rs lattice \cite{Galasp1,Galasp2,Galasp3}. In Fig.
\ref{fig1} we present both the zb and rs structures which are
similar to the cubic lattice of the Heusler compounds with two
voids per unit cell. Evidence of the growth of such
nanosctructures has been provided by Liu et al who have reported
successful self-assembly growth of ultrathin CaN in the rocksalt
structure on top of Cu(001) \cite{Liu08}. Finally we have to note
that materials containing C or N seem to be more promising for
applications since the Hund energy for the light atoms in the
second row of the periodic table is similar to the Hund energy of
the 3d transition metal atoms.

Slater and Pauling had shown in two pioneering papers that in the
case of binary magnetic compounds when we add one valence electron
in the compound this occupies spin-down states only and the total
spin magnetic moment decreases by about 1 $\mu_B$
\cite{Slater,Pauling}. Interestingly a similar behavior can be
also found in half-metallic magnets as confirmed by
first-principles (ab-initio) electronic structure calculations. In
the half-metallic alloys the spin-down band-structure is fixed;
the number of spin-down occupied bands and their character does
not change among the half-metallic members of the same family of
compounds. The extra valence electron now occupies exclusively
spin-up states increasing the total spin magnetic moment by about
1 $\mu_B$. It was shown that in the case of the semi-Heusler
compounds like NiMnSb the total spin magnetic in the unit cell,
$M_t$ scales, as a function of the total number of valence
electrons, $Z_t$, following the relation $M_t=Z_t-18$
\cite{Galanakis02}, while in the case of the L$2_1$ full-Heuslers
this relation becomes $M_t=Z_t-24$ \cite{Galanakis02b}. Moreover
for the TM pnictides and chalcogenides as well as the
d$^0$-ferromagnets the corresponding
 rule takes the form   $M_t=Z_t-8$ although its origin is different in the two families of
 compounds \cite{GalaZB,Galasp1,Galasp2}
These Slater-Pauling (SP) rules connect the electronic properties
(appearance of the half-metallic behavior) directly to the
magnetic properties (total spin magnetic moments) and thus offer a
powerful tool to the study of half-metallic compounds since (i)
magnetic measurements can be used to confirm the half-metallic
character of a compound, and (ii) simple valence electrons
counting can predefine the magnetic properties of a half-metal.
The aim of the present Short Review is to provide an overview of
the origin of the SP rule in all mentioned families of
half-metallic magnets as derived from electronic structure
calculations based on the density functional theory.

\begin{figure*}
\begin{center}
\includegraphics[width=\textwidth]{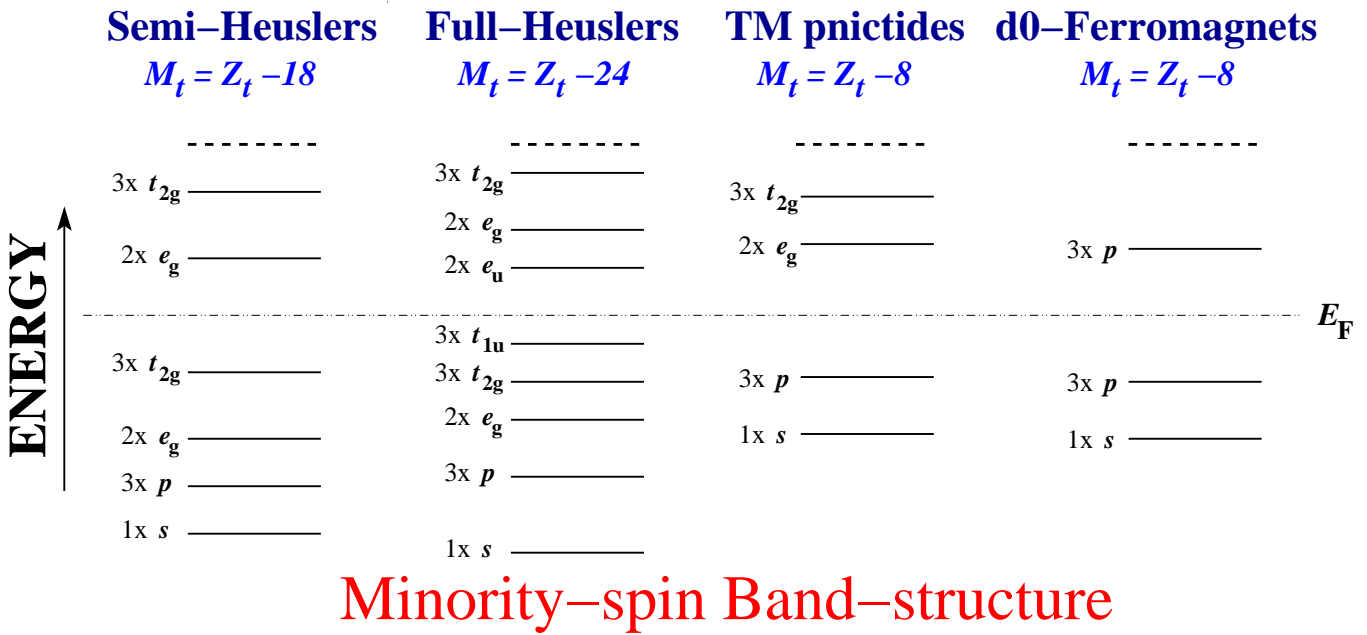}
\end{center}
\caption{Schematic representation of the energy levels of the
minority-spin electronic band structure for all four families of
half-metallic magnets under study. Below the Fermi level are
located the occupied states. Numbers in front of orbitals denote
the corresponding degeneracy. For the definition of the orbitals
see text.} \label{fig2}
\end{figure*}

\section{Heusler compounds}\label{sec2}

\begin{figure}
\begin{center}
\includegraphics[width=\columnwidth]{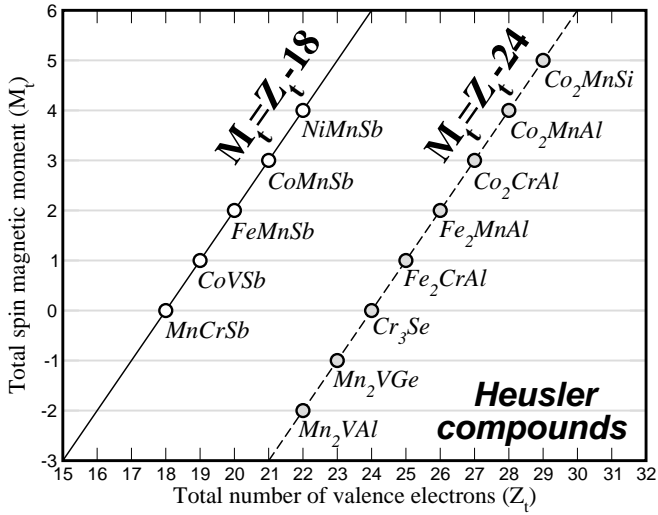}
\end{center}
\caption{Calculated total spin magnetic moments, $M_t$, in $\mu_B$
as a function of the total number of valence electrons, $Z_t$, in
the unit cell for selected representative semi- and full-Heusler
compounds.  The two lines represent the two variants of the
Slater-Pauling rule followed by the two families of Heuslers.}
\label{fig3}
\end{figure}

The most studied half-metallic magnets are the Heusler compounds.
We will start our discussion from the  case of semi-Heuslers, like
NiMnSb, since it was the first to be predicted to be a half-metal
\cite{deGroot}. The role of the \textit{sp} element is to provide
in the spin-down electronic band structure a single \textit{s} and
a triple-degenerated \textit{p} band deep in energy; they are
located below the \emph{d}-states and accommodate \emph{d}-charge
from the transition metal atoms. In Fig. \ref{fig2} we present
schematically the character of the bands and their degeneracy in
the spin-down band structure. The bands below the Fermi level are
occupied while the ones above it are empty.
 The \emph{d}-orbitals of the two transition metal atoms
hybridize strongly creating five occupied bonding and five
unoccupied antibonding \emph{d}-states in the spin-down band
structure \cite{Galanakis02}. Each set of five
occupied(unoccupied) $d$-hybrids contains the double degenerated
$e_g$ and the triple degenerated $t_{2g}$ states. Note that the
notations $e_g$ and $t_{2g}$ are, strictly speaking, valid only
for states at the center of the Brillouin zone; however, the
energy bands formed by them are energetically rather separated,
and therefore this notation can be used to describe the different
bands. As a result there are in total exactly nine occupied
spin-down states and the SP relation is $M_t=Z_t-18$, where $M_t$
is the total spin magnetic moment in $\mu_B$ and $Z_t$ the total
number of valence electrons \cite{Galanakis02}. The number of the
occupied spin-down states is always kept equal to nine for all
half-metallic semi-Heusler compounds (the Fermi level is fixed
within the spin-down energy gap) and does not depend on the
chemical species  of the constituent atoms and thus the total spin
magnetic moment depends only on the number of valence electrons
and not on the specific chemical type of the half-metal,
\textit{e.g.} FeMnSb and CoCrSb have both 20 valence electrons and
a total spin magnetic moment of 2 $\mu_B$. When the total spin
magnetic moment is positive and thus are more than 18 valence
electrons in the unit cell, in the spin-up band structure all nine
bonding \emph{s}-, \emph{p}- and \emph{d}-states are occupied, as
in the spin-down band, and the extra charge occupies the
antibonding spin-up states \cite{Galanakis02}. In Fig. \ref{fig3}
we have gathered the calculated total spin magnetic moments versus
the total number of valence electrons for some selected
representative semi Heusler compounds. A special case is MnCrSb,
which has a zero total spin magnetic moment but Mn and Cr atoms
posses large antiparallel spin magnetic moments \cite{Leuken}.
These compounds are known as half-metallic antiferromagnets and
are of technological importance since they create in-principle
zero external fields and thus minimal energy losses in devices.

In the case of the half-metallic L$2_1$ full-Heuslers the origin
of the SP rule is more complicated due to the more complicated
hybridization effects between the \emph{d}-orbitals
\cite{Galanakis02b}. The \textit{sp} element provides in the
spin-down electronic band structure a single \textit{s} and a
triple-degenerated \textit{p} band deep in energy as for the
semi-Heusler compounds. With respect to the $d$-orbtials, one has
first to consider the interaction between the X elements. Although
the symmetry of the L$2_1$ lattice is the tetrahedral one, the X
elements themselves, if we neglect the Y and Z atoms, form a
simple cubic lattice and sit at sites of octahedral symmetry
\cite{Galanakis02b}.  The \emph{d}-orbitals of the neighboring X
atoms hybridize creating five bonding \emph{d}-states, which after
hybridize with the \emph{d}-orbitals of the Y atoms creating five
occupied and five unoccupied \emph{d}-hybrids, and five
non-bonding \emph{d}-hybrids of octahedral symmetry (the
triple-degenerated $t_{1u}$ and double-degenerated $e_u$ states).
These non-bonding hybrids cannot couple with the orbitals of the
neighboring atoms, since they do not obey the tetrahedral
symmetry, and only the $t_{1u}$ are occupied leading to a total of
12 occupied spin-down states  and the SP relation is now
$M_t=Z_t-24$ \cite{Galanakis02b}.  In the case of full-Heusler
compounds when $Z_t>24$ the spin-up non-bonding $e_u$ states are
the first to be occupied followed by the antibonding states, while
when $Z_t<24$ the Fermi level crosses either the spin-up
non-bonding $t_{1u}$ states or the spin-up bonding \emph{d}-states
\cite{Galanakis02b}. In Fig. \ref{fig3} we have gathered some
representative cases of calculated half-metallic full-Heusler
compounds. The total spin magnetic moment can be: (i) zero as for
Cr$_3$Se, which is a half-metallic antiferromagnet \cite{Cr3Se},
(ii) positive when $Z_t>24$  going up to a maximum value of 5
$\mu_B$ for Co$_2$MnSi (it has been reported that electronic
correlation restore half-metallicity in Co$_2$FeSi which has a
total spin magnetic moment of 6 $\mu_B$ \cite{Co2FeSi}), or (iii)
negative as for the half-metallic ferrimagnetic Mn$_2$VGe and
Mn$_2$VAl compounds where the  V spin moment is antiparallel to
the spin moment of the Mn atoms \cite{Mn2VZ}.

\begin{figure}
\vskip 1cm
\begin{center}
\includegraphics[width=\columnwidth]{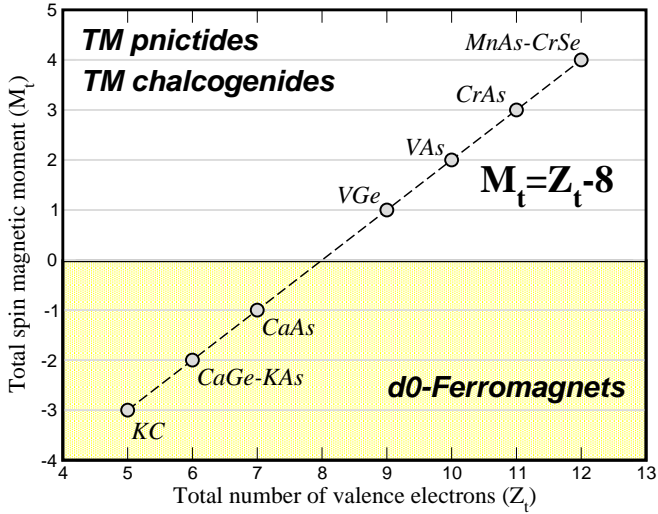}
\end{center}
\caption{Similar to Fig. \ref{fig3} for the binary half-metallic
magnets under study. Both transition-metal pnictides
(chalcogenides) and d$^0$-ferromagnets follow the same
Slater-Pauling rule despite the different origin of the rule (see
Fig. \ref{fig2}).} \label{fig4}
\end{figure}

\section{Binary compounds}\label{sec3}

\subsection{TM pnictides \& chalcogenides}

The principal mechanism leading to the appearance of the
half-metallic gap in zincblende compounds of transition metals
(TM) with $sp$ elements is the hybridization of the $d$
wavefunctions of the TM atom with the $p$ wavefunctions of the
$sp$ atom \cite{GalaZB}. This \emph{$p$-$d$} hybridization is
dictated by the tetrahedral environment, where each atom is
surrounded by four atoms of the other species as shown in Fig.
\ref{fig1}. In particular, in the presence of tetrahedral symmetry
the $d$-states split in two irreducible subspaces: the $t_{2g}$
threefold-degenerate  subspace, consisting of the $d_{xy}$,
$d_{yz}$ and $d_{xz}$ states, and the twofold-degenerate $e_{g}$
subspace, which includes the $d_{x^2-y^2}$ and $d_{z^2}$ states.
Only states of the former subspace can hybridize with the $p$
states of the $sp$-atom neighbors, forming bonding and antibonding
hybrids. In the spin-down band structure the bonding states are
mainly of $p$ character while the antibonding states of $t_{2g}$
character.  The $e_{g}$ states of the TM atom, on the contrary,
remain rather non-bonding. The situation is shown schematically in
Fig. \ref{fig2}  for the spin-down band structure. Below the Fermi
level there is a single-degenerated $s$-band due to the $sp$ atom
followed by the bonding $p-d$ hybrids which are mainly of $p$
character with a small $t_{2g}$ admixture. Above the Fermi level
are the non-bonding TM $e_g$ states followed by the antibonding
$p-d$ hybrids which are mainly of $t_{2g}$ character.

Thus for the transition-metal pncitides and chalcogenides there
are in total four occupied states in the spin-down band-structure
and the SP rule becomes  $M_t=Z_t-8$ \cite{GalaZB}. In Fig.
\ref{fig4} we have gathered  the calculated total spin magnetic
moment for some transition metal pnictides and chalcogenides
versus the total number  of valence electrons. The total spin
magnetic moment can vary from 1 $\mu_B$ for VGe which has 9
valence electrons per unit cell up to 4 $\mu_B$ for MnAs and CrSe
which have 12 valence electrons. Compounds with 13 valence
electrons like MnSe prefer to be antiferromagnets that
half-metallic ferromagnets since in the latter case all
antibonding $p-d$ hybrids in the spin-up band-structure should be
occupied which is energetically unfavorable \cite{GalaZB}.

\subsection{d$^0$-Ferromagnets}

In the case of d$^0$-ferromagnets like CaAs the SP rule is
identical to the TM pnictides and chalcogenides if we consider
that the energy gap is located in the spin-down band structure; in
most references like \cite{Galasp1,Galasp2,Galasp3} the energy gap
is considered to be located in the spin-up band structure and the
SP rule is $M_t=8-Z_t$. Its origin is different than the TM
pnictides, discussed above since the d$^0$-ferromagnets are made
up of two $sp$ atoms and thus no $p-d$ hybridization is present.
In the spin-down band-structure the $s$  and $p$ orbitals of the
two $sp$ atoms hyrbidize creating bonding and antibonding hybrids.
In the spin-down band-structure, as shown in Fig. \ref{fig2}, the
bonding $s$ and $p$ states are occupied while the antibonding ones
are empty leading to a total of 4 occupied spin-down bands. In the
spin-up band structure the bonding $s$ state is occupied and the
Fermi level crosses the bonding $p$-states since we have less than
8 valence electrons in the unit cell (compounds with 8 valence
electrons like CaS are semiconductors). In Fig. \ref{fig4} we have
gathered the calculated total spin magnetic moments for some
representative d$^0$-ferromagnets and the total moment ranges from
-3 $\mu_B$ for KC which has just 5 valence electrons per unit cell
up to -1 $\mu_B$ for CaAs which has 7 valence electrons in the
unit cell. Finally we should note that the arguments just
presented  are valid for both the rocksalt (which is the most
stable structure) and the zincblende structure since as presented
in Fig. \ref{fig1} both obey the tetrahedral symmetry and similar
symmetry arguments are valid.

\section{Conclusions}\label{sec4}

Half-metallic magnets are a special class of materials which has
attracted considerable attention in spintronics/magnetoelectronics
due the predicted perfect spin-polarization of the electrons at
the Fermi level. In the case of Heusler and binary half-metallic
compounds, it is possible to formulate Slater-Pauling rules which
connect directly the electronic properties (number of valence
electrons in the unit cell) to the magnetic properties (total spin
magnetic moment in the unit cell). The Fermi level is fixed within
the spin-down energy gap and extra electrons accommodate
exclusively spin-up states. These rules are a strong tool in the
study of the half-metallic magnets since: (i) magnetic
measurements can be employed to confirm the half-metallic
character of samples, and (ii) the chemical formula of a
half-metallic compound predefines its total spin magnetic moment.
The origin of the Slater-Pauling rule differs in the various
families of compounds under study and for each case the possible
hybridization of the orbitals with respect to both the local and
the crystal's  symmetry should be taken into account. In
semi-Heusler compounds, like NiMnSb, the hybridization of the
$d$-orbitals leads to a $M_t=Z_t-18$ relation, where $M_t$ the
total spin magnetic  moment in $\mu_B$ and $Z_t$ the total number
of valence electrons in the unit cell. In the case of
full-Heuslers, like Co$_2$MnSi, the existence of $d$-hybrids of
octahedral symmetry located exclusively at the Co sites leads to a
more complicated hybridization effect and a $M_t=Z_t-24$ rule.
Finally for both type of binary compounds the Slater-Pauling rule
is $M_t=Z_t-8$ but its origin is the $p-d$ hybridization in the
transition-metal pnictides and chalcogenides and the $p-p$
hybridization in the d$^0$-ferromagnets.

\end{document}